\documentclass[usenatbib,usegraphicx]{mn2e}
\usepackage{apjfonts}
\usepackage{journals}

\begin{document}

\title[Probing the structure of white dwarfs]{Probing the core 
       and envelope structure of DBV white dwarfs}

\author[Metcalfe, Montgomery, \& Kawaler]{T. S. Metcalfe,$^1$
        M. H. Montgomery,$^2$ \& S. D. Kawaler$^3$ \\
$^1$ Harvard--Smithsonian Centre for Astrophysics,
     60 Garden Street, Cambridge, MA 02138 \\
$^2$ Institute of Astronomy, University of Cambridge,
     Madingley Road, Cambridge CB3 0HA, UK \\
$^3$ Department of Physics and Astronomy, 
     Iowa State University, Ames, IA 50011}

\maketitle

\begin{abstract}
We investigate the global pulsation properties of DBV white dwarf models
that include both the double-layered envelope structure expected from
time-dependent diffusion calculations, as well as a non-uniform C/O core
expected from prior nuclear burning. We compare these models to otherwise
identical models containing a pure C core to determine whether the
addition of core structure leads to any significant improvement. Our
double-layered envelope model fit to GD~358 that includes an adjustable
C/O core is significantly better than our pure C core fit ($7\sigma$
improvement). We find a comparable improvement from fits to a second DBV
star, CBS~114, though the values of the derived parameters may be more
difficult to reconcile with stellar evolution theory. We find that our
models are systematically cooler by 1,900 K relative to the similar models
of \citet{fb02}. Although a fit to their model reproduces the mass and
envelope structure almost exactly, we are unable to reproduce the absolute
quality of their fit to GD~358. Differences between the constitutive
physics employed by the two models may account for both the temperature
offset and the period residuals.
\end{abstract}


\section{INTRODUCTION}

The vast majority of what we can learn about stars is derived from
observations of the thin outermost skins of their surface layers. For the
most part, we are left to infer the properties of the interior based upon
our current best understanding of the constitutive physics. There are a
number of exceptions to this general rule, including observations of
supernovae explosions and interacting binary stars, where the deeper
layers of the interior are either ejected from the system or are gradually
exposed as mass is transferred from one star to the other, respectively.
However, both of these processes are disruptive to the stellar interior,
making inferences about the original structure ambiguous at some level.

Pulsating stars represent the best opportunity for probing stellar
interiors while they are still intact. The most dramatic example is the
Sun, where observations of light and radial velocity variations across the
visible surface have led to the identification of thousands of unique
pulsation modes, each sampling the solar interior in a slightly different
and complementary way. These observations have led to such precise
constraints on the standard solar model that the inverted radial profile
of the sound speed, for example, now agrees to better than a few parts per
thousand over 90 per cent of the solar radius \citep{jcd02}.

If we were to move the Sun to the distance of even the nearest star, most
of the pulsation modes that we now know to be present would be rendered
undetectable. We would lose nearly all of our spatial resolution across
the disc of the star, and only those modes of the lowest spherical degree
($\ell\la3$) would produce significant variations in the total integrated
light or the spectral line profiles. This would reduce the number of
detectable modes from thousands to merely dozens, leading to a
corresponding reduction in the ability of the observations to constrain
the internal structure \citep[e.g., see][]{kje99}. Even so, such data
would still allow us to determine the global properties of the star and to
probe the gross internal composition and structure, providing valuable
independent tests of stellar evolution theory.

Pulsations in white dwarfs were first discovered in the cooler DA stars by
\citet{lan68}, and later in the hotter PG~1159 stars \citep{mcg79} and DB
stars \citep{win82}. These differ from the solar oscillations in that they
are non-radial $g$-modes instead of $p$-modes; the restoring force is
gravity rather than pressure, and so the most important physical quantity
is the buoyancy frequency, not the acoustic frequency. White dwarf stars
are the end-points of stellar evolution for all stars with masses below
what is necessary to produce elements much heavier than carbon and oxygen.
Their interior structure contains a record of the physical processes that
operate during the later stages in the lives of most stars, so there is a
potential wealth of information encoded in their pulsation frequencies.

Much of the observational data for white dwarf asteroseismology has come
from an international collaboration known as the Whole Earth Telescope
\citep[WET;][]{nat90}. This network of small telescopes situated around
the globe obtains the long time-series of photometric measurements that
are often necessary to unambiguously identify and resolve dozens of unique
pulsation modes. One of the most notable successes of this collaboration
was an observing campaign in 1990 on the brightest known DB variable,
GD~358 \citep{win94}. Among many interesting results, these observations
established the simultaneous presence of 11 low-degree ($\ell$=1, $m$=0)  
modes of consecutive radial overtone ($k$=8--18) with periods in the range
400--800 seconds \citep{bw94}.

\begin{figure}
\centering
\includegraphics[width=\columnwidth]{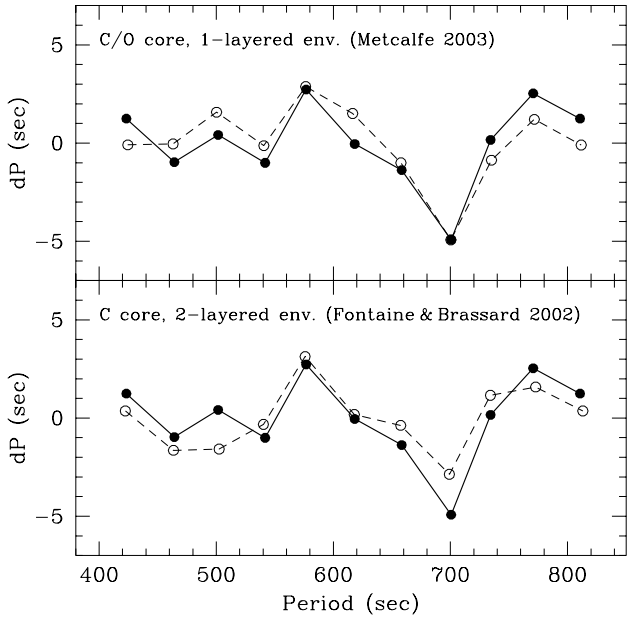}
\caption[MD570L_fig1.eps]{The pulsation periods observed in GD~358 (solid 
points) plotted against their deviations from the mean period spacing (dP),
along with two physically distinct model fits (open points).
The fit of \citet{met03} has extra structure in the core (top panel),
while the fit of \citet{fb02} has extra structure in the envelope (bottom
panel).\label{fig1}}
\end{figure}

The theoretical interpretation of these data has grown gradually more
sophisticated as our computational capabilities have expanded over the
decade following the observations. The most recent analysis produced a
model that leads to a root-mean-square difference between the observed and
calculated periods of only $\sim$1 second \citep{met03}. However, a
comparable match to the same observations was recently published by
\citet{fb02}, who used a completely independent model with an internal
structure that was physically distinct from that assumed by
\citeauthor{met03} (see Fig.~\ref{fig1}). Essentially, \citeauthor{fb02}'s
model contains additional structure in the envelope, while
\citeauthor{met03}'s model has extra structure in the core.

While this duality in model-fits is now understood to be due to an
inherent symmetry in the way the pulsations sample the interior
\citep{mmw03}, there is good reason to believe that the physical basis of
each model is sound, but that neither represents a complete description of
the true interior structure. In this paper, we attempt to bridge the gap
between these two models by including the essential elements of each into
a `hybrid' model that contains both the double-layered envelope structure
expected from time-dependent diffusion calculations, and an adjustable
carbon/oxygen (C/O) core. In \S\ref{SEC2} we outline the physical basis of
our model parametrizations, and in \S\ref{SEC3} we present the results of
our model-to-model and model-to-data comparisons. We conclude in
\S\ref{SEC4} with a brief discussion of the promise of asteroseismology.

\begin{figure}
\centering
\includegraphics[width=\columnwidth]{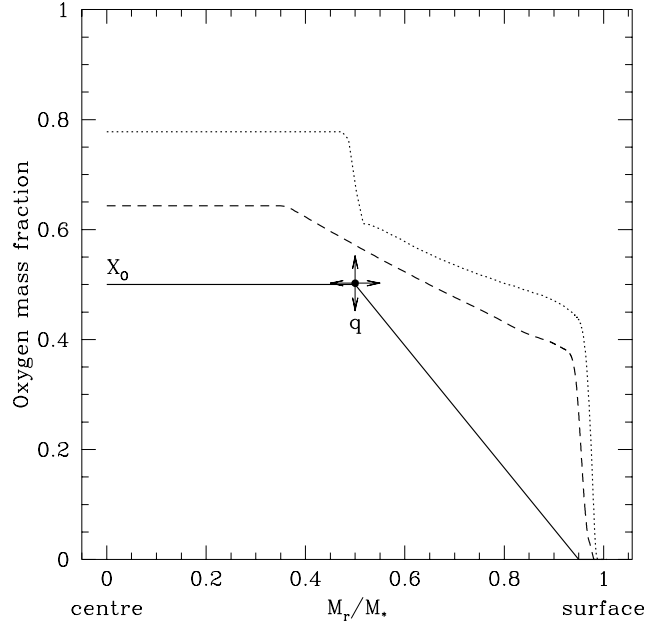}
\caption[MD570L_fig2.eps]{Theoretical white dwarf internal oxygen profiles 
from the calculations of \citet[][dotted]{sal97} and from
\citet[][dashed]{alt02}, along with the simplified generic profiles that
we adopt for the models in this paper (solid) in which both the central
oxygen mass fraction ($X_{\rm O}$, vertical axis) and the fractional mass
of the initial break from a uniform C/O mixture ($q$, horizontal axis) are
adjustable parameters.\label{fig2}}
\end{figure}

\section{THEORETICAL MODELS\label{SEC2}}

\subsection{Adjustable C/O Cores}

The generic shape of a theoretical white dwarf internal oxygen profile is
set by the nuclear and mixing processes that occur during its formation in
the core of a red giant. The detailed shape is less certain, since it
depends on the specific physical and numerical treatments utilized by the
adopted model. Many models agree that the inner $\sim$0.5 in fractional
mass should contain an approximately uniform C/O mixture
\citep{sal97,alt02}.  The precise extent of this region is determined by
the maximum size of the central convective core (and by the amount of
convective overshooting, if it occurs) during helium burning. The C/O
ratio in this uniform region is primarily set by the rate of the
$^{12}{\rm C}(\alpha,\gamma)^{16}{\rm O}$ reaction. Outside of the uniform
C/O core the exact manner in which the oxygen mass fraction drops to zero
is uncertain, since it depends on the adopted recipe for semiconvection
during core helium burning and on the physical conditions in the models
during helium shell burning. However, the location and slope of the
initial break from a constant C/O ratio seems to be the most important
feature from an asteroseismological standpoint.

Our parametrization for the C/O core is identical to that used in the
earlier model fitting of \citet{met03}, \citet{msw02}, and \citet{mwc01}.
We fix the oxygen mass fraction to its central value ($X_{\rm O}$) out to
some fractional mass ($q$) where it then decreases linearly in mass to
zero oxygen at the 0.95 fractional mass point (see Fig.~\ref{fig2}). This
form for the profile was chosen to facilitate comparison with the earlier
work of \citet{bww93}.

Despite this simplification, \citet{met03} obtained fits to two different
pulsating DB white dwarfs that independently led to implied rates for the
$^{12}{\rm C}(\alpha,\gamma)^{16}{\rm O}$ reaction that were consistent
both with each other and with recent extrapolations from high energy
laboratory measurements. While these fits are the best evidence to date
that our models really are sensitive to the core composition and structure
sampled by the white dwarf pulsations, they assumed a single-layered
structure for the transition between the C/O core and the pure He
envelope. If there is an evolutionary connection between the DB stars and
their hotter cousins, the PG~1159 stars, then this single-layered
structure may not be an adequate representation of the actual white dwarf
envelopes.

\begin{figure}
\centering
\includegraphics[width=\columnwidth]{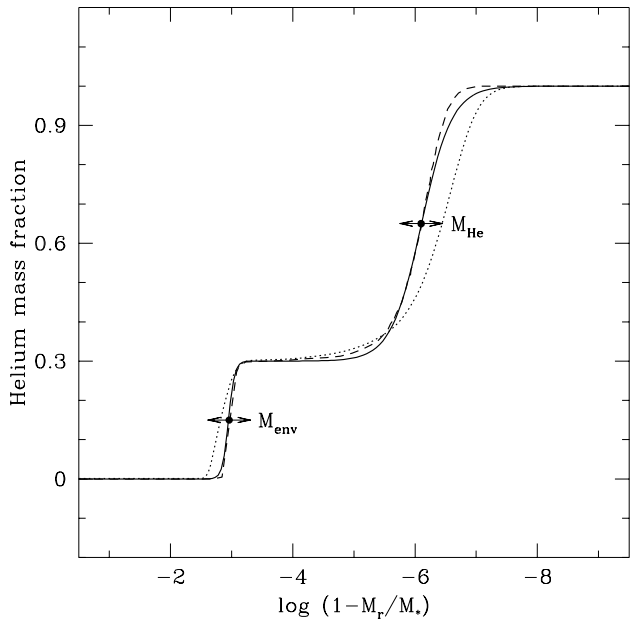}
\caption[MD570L_fig3.eps]{Internal He profiles from the calculations of
\citet[][dotted]{dk95} and from \citet[][dashed]{fb02}, along with a
profile from the parametrization that we adopt for the models in this
paper (solid). The locations of the two chemical transitions (points) are
adjustable parameters, but the shapes are fixed.\label{fig3}}
\end{figure}

\subsection{Double-Layered Envelopes}

\citet{dk95} were the first to demonstrate the plausibility of an
evolutionary connection between PG~1159 and DB stars. They created an
initial model based on the pulsation data for PG~1159 \citep{win91}, which
included a thick envelope [$\log(M_{\rm env}/M_*)\sim-2.5$] containing a
uniform mixture of 30\% He, 35\% C, and 35\% O above the C/O core. During
the evolution of this model to lower temperatures, they included
time-dependent diffusive processes so they could monitor the chemical and
gravitational separation of the elements -- essentially He `floating' to
the surface. By the time the model had reached the temperature range of
the DB variables, the transition between the pure He surface and the
still-uniform He/C/O mantle was located near $\log(M_{\rm
He}/M_*)\sim-5.5$, leading to a structure that looked very similar to the
outer regions of the fit of \citet{bw94}. The pulsation periods of this
model were a good match to the WET observations of GD~358, even though no
formal attempt was made to fit them.

More recently, \citet{fb02} calculated a large grid of similar models that
included this double-layered envelope structure in an attempt to match the
pulsation periods of GD~358 in detail. Using models that contained a pure
C core, they were able to find a match with root-mean-square period
residuals nearly as low as the fit of \citet{met03}. They also attempted
similar fits using models with cores of pure O and a uniform C/O mixture,
resulting in models that were only `marginally worse'. They speculated
that the 11 periods observed in GD~358 may not contain enough information
to measure the core and envelope structure simultaneously.

To investigate this possibility more thoroughly, we replaced the
single-layered envelopes in our models with a parametrization of the
double-layered structure used by \citet{fb02}. This allows us to specify
the locations in the envelope where the He mass fraction reaches 0.15 (the
middle of the transition between the C/O core and the He/C mantle) and
0.65 (the middle of the transition between the mantle and the pure He
surface)\footnotemark[1]. The He profiles from several different models
are shown in Fig.~\ref{fig3}. Although there are slight differences
between the profile from our parameterization compared to that of
\citet{fb02}, we have verified that modifications to the detailed {\it
shape} of the profile comparable to these differences lead to only
marginal improvements in the final fits. Note that the two layers are 
theoretically expected to merge into one at low values of $T_{\rm eff}$ 
\citep{ga02}, leading to structures that have effectively been explored by 
the single-layered global model-fitting of \citet{mwc01} and \citet{hmw02}
for GD~358 and CBS~114 respectively.

  \footnotetext[1]{\citeauthor{fb02} define the location of the outer 
  transition to be where the He mass fraction reaches 0.50, which is 
  systematically deeper by $\sim$0.3 dex relative to our definition.}

\begin{table*}
\centering
\begin{minipage}{100mm}
\caption{Double-layered envelope models, with and without an adjustable C/O 
core, fit to the model periods of \citet{fb02} and to the periods observed 
in the pulsating DB white dwarf stars GD~358 and CBS~114.\label{tab1}}
\begin{tabular}{@{}lrrcrrcrr@{}}
\hline
& \multicolumn{2}{c}{F\&B Model} &
& \multicolumn{2}{c}{GD~358}  &
& \multicolumn{2}{c}{CBS~114} \\ 
\cline{2-3} \cline{5-6} \cline{8-9} 
Parameter                      
& pure C  & C/O\footnote{Single-layered envelope fit from \citet{mmw03}} &
                               & pure C           & C/O           &     
                               & pure C           & C/O           \\
\hline
$T_{\rm eff}$~(K)$\dotfill$    & 22,900           & 22,900        &
                               & 21,500           & 20,800        & 
                               & 24,500           & 25,800        \\
$M_*\ (M_{\odot})\dotfill$     & 0.625            & 0.660         &
                               & 0.665            & 0.670         &
                               & 0.650            & 0.615         \\
$\log(M_{\rm env}/M_*)\ldots$  & $-$2.94          & $-$2.00       &
                               & $-$3.10          & $-$2.68       &
                               & $-$3.96          & $-$3.78       \\
$\log(M_{\rm He}/M_*)\dotfill$ & $-$6.10\footnote{By \citeauthor{fb02}'s 
definition, $\log(M_{\rm He}/M_*)=-5.77$}         &$\cdots$       &
                               & $-$6.22          & $-$5.22       &
                               & $-$5.72          & $-$5.74       \\
$X_{\rm O}\dotfill$            & $\cdots$         & 0.67          &
                               & $\cdots$         & 0.75          &
                               & $\cdots$         & 0.71          \\
$q\dotfill$                    & $\cdots$         & 0.48          &
                               & $\cdots$         & 0.49          &
                               & $\cdots$         & 0.56          \\
$\sigma_{\rm P}$~(s)$\dotfill$ & 1.63             & 1.26          &
                               & 2.17             & 1.52          &
                               & 0.82             & 0.27          \\
\hline
\end{tabular}
\vspace*{-18pt}
\end{minipage}
\vspace*{12pt}
\end{table*}

\vspace*{-12pt}
\section{NEW MODEL FITTING\label{SEC3}}

To get a sense of the relative improvement that might be possible by
including an adjustable C/O core in our models, we performed two sets of
fits to the pulsation data. In both cases, we used a genetic algorithm
\citep[see][]{mc03} to minimize the root-mean-square residuals between the
observed and calculated periods ($\sigma_{\rm P}$) for models with
effective temperatures ($T_{\rm eff}$) between 20,000 and 30,000 K, and
total stellar masses ($M_*$) between 0.45 and 0.95 $M_{\odot}$. We allowed
the location of the inner He transition (between the core and the mixed
He/C mantle) to assume values of $\log(M_{\rm env}/M_*)$ between $-2.0$ and
$-4.0$, and for the outer He transition (between the mixed He/C mantle and
the pure He surface) the value of $\log(M_{\rm He}/M_*)$ was in the range
$-5.0$ to $-7.0$. For the first set of fits we fixed the core composition to 
be pure C. For the second set we included an adjustable C/O core, allowing
the central oxygen mass fraction ($X_{\rm O}$) to vary between 0.0 and 1.0,
with the break from a uniform C/O mixture beginning at a fractional mass
($q$) between 0.10 and 0.85.

We performed these fits on three sets of pulsation data. To explore the
systematic differences between our own models and those of \citet{fb02},
we fit the 11 model pulsation periods of their fit to GD~358, which had a
pure C core. For our adjustable C/O fit to these periods, we used a
single-layered envelope model as part of an investigation of the symmetry
between these two types of models \citep{mmw03}. The second set of data
came from the 1990 WET campaign on GD~358 \citep{win94}, which is the same
set of 11 periods fit by \citet{fb02}. To investigate whether the
double-layered envelope models could produce fits to a second DBV star
that were physically consistent with time-dependent diffusion theory, our
third set of data came from the slightly hotter white dwarf CBS~114, which
exhibits 7 independent pulsation periods \citep{hmw02}. In all cases we
restricted our models to calculate the $\ell$=1, $m$=0 modes, as suggested
by the observations. The results of our two fits to each of these three
data sets are shown in Table \ref{tab1}.

\vspace*{-12pt}
\subsection{Carbon Core Models}

From our pure C core fit to the model periods of \citet[][their table
1]{fb02} we retrieved the same mass and virtually identical locations for
the two He transitions in the envelope, but our temperature was
systematically low by 1,900 K relative to their value (24,800 K). Because
the internal structure of the two models is very similar, this temperature
offset is most likely due to differences between the constitutive physics
(equations of state, opacities, etc.) as well as the convective
prescription employed by each.  In a study of the influence of various
updates to the constitutive physics on the pulsational properties of their
models, \citet{fb94} found that the OPAL radiative opacities \citep{ir93}
for pure He and C were systematically lower than the older LAO data
\citep{hue77}, and that models using the higher opacity data could mimic
hotter models with otherwise identical parameters. This is consistent with
the results of our fit, since our models use the more opaque LAO data,
while \citeauthor{fb02} use the more recent OPAL data \citep{fbb01}.
However, our preliminary attempts to use the OPAL data in our own code
suggest that this may only account for roughly half of the temperature
difference between the models. Other possible contributions include
differences between our equations of state -- both of our models use
tables derived from \citet{lv75} and \citet{fgv77} for parts of the
interior and the envelope respectively, but \citeauthor{fb02}'s models
include additional unpublished modifications \citep{fbb01}. More work will
be required to document these differences and to evaluate their influence
on the effective temperature scales of our models.

Our fit to the observed pulsation periods of GD~358 has a higher mass and
a lower temperature than the fit of \citet{fb02}, but the locations of the
two He transitions are both within their $1\sigma$ uncertainties. The most
striking difference between our fits is that while \citeauthor{fb02} were
able to match the periods observed in GD~358 with residuals of only
$\sigma_{\rm P}=1.30~s$, our optimal fit has $\sigma_{\rm P}=2.17~s$ (a
difference nearly 30 times larger than the observational uncertainty,
$\sigma_{\rm obs}\sim0.03~s$). Again, differences between the constitutive
physics employed by the two models may account for at least part of this
difference in the {\it absolute} quality of the two fits -- using the OPAL
radiative opacities with our models led to a significant decrease in the
residuals (e.g., from 1.63 to 1.45~$s$ in the case of our fit to
\citeauthor{fb02}'s model). Despite the unresolved differences between the
constitutive physics, we can still use the {\it relative} quality of our
own fits to determine whether or not the addition of extra free parameters
in a given model lead to significant improvements.

Our double-layered envelope fit to CBS~114 is only marginally
(2$\sigma_{\rm obs}$) better than the pure C core single-layered envelope
model fit of \citet{hmw02}, compared to what is expected from the addition
of an extra parameter. Our values for the mass, the temperature, and the
(inner) He transition are indistinguishable from the single-layered fit.
Although the derived temperature for CBS~114 is higher than for GD~358, as
expected from the spectroscopic measurements of \cite{bea99}, the pure He
surface layer for this fit is {\it thicker} than for GD~358 -- just the
opposite of what time-dependent diffusion calculations would lead us to
expect for models with similar masses.

\vspace*{-12pt}
\subsection{C/O Core Models}

When we used an adjustable C/O core and a {\it single-layered} envelope to
fit the model periods of \citet{fb02}, it led to a significant improvement
to our match even though the interior structure of our model was
dramatically different from the source model. Our derived value of $q$
corresponds {\it exactly} to the true location of the outer He transition
reflected through the core/envelope symmetry described by \citet{mmw03}.
Essentially our fit confirms empirically that such a symmetry exists, and
that it is possible to fit real structure in the envelopes by assuming
structure at the corresponding symmetric location in the core.

Our fit to GD~358 leads to inferred locations for all chemical transition
zones that are distinct from each other with respect to the core/envelope
symmetry. Statistically, the addition of an adjustable C/O core to the
double-layered envelope fit leads us to expect the residuals to decrease
to $1.74~s$ just from the addition of the two extra parameters. In fact
the residuals of our C/O fit are reduced to $1.52~s$, a 7$\sigma_{\rm
obs}$ improvement {\it beyond what is expected}. In this model, the C/O
core is effectively fitting the same mode trapping structure attributed to
the outer He transition in \citeauthor{fb02}'s fit. Our outer He
transition, centered at $\log(M_{\rm He}/M_*) = -5.22$, is distinct since
no structure is expected at the corresponding symmetric location in the
core, which is near $M_{\rm r}/M_* \sim 0.6$ \citep[see][their
fig.~3]{mmw03}.

The improvement in the fit to CBS~114 is even more substantial. While the
derived locations of the two He transitions do not change by much between
the pure C and the C/O fits, the addition of structure in the core leads
to an improvement in the fit of nearly 12$\sigma_{\rm obs}$ {\it beyond what
is expected from the extra parameters}. As with GD~358, the locations of
the core and outer envelope structure are distinct from one another, from
the standpoint of the symmetry inherent in the models. Interestingly, the
conflict between the higher temperature and the thickness of the surface
He layer disappears for the C/O fits. However, both fits to CBS~114 lead
to an unusually small total envelope mass relative to the range expected
[$\log(M_{\rm env}/M_*)\simeq -2.0$ to $-3.0$] from simulations of carbon
dredge-up in DQ stars \citep[see][]{pel86,fb02}. In addition, the masses
of the C/O fits are in conflict with the spectroscopic measurements of
\citet{bea99}.

\vspace*{-12pt}
\section{DISCUSSION\label{SEC4}}

Our global exploration of models containing double-layered envelope
structure make it clear that the addition of an adjustable C/O core leads
to significantly better fits to the observations, relative to pure C core
models. This is reassuring, since there are sound physical reasons that
lead us to expect composition transition zones in both the cores and the
envelopes of real white dwarf stars.

The {\it absolute} quality of our fits to GD~358 are generally worse than
the fits of \citet{fb02} or \cite{met03}. If white dwarf envelopes really
do contain a double-layered structure, this implies that we will need to
know the detailed shapes of the composition transition zones in both the
core and the envelope before better fits can be found.

The fits to CBS~114 support our general conclusion that the models are
sensitive to structure in {\it both} the core and the envelope, but only
the C/O fit is consistent with the expectations from time-dependent
diffusion theory (thinner surface He layers for hotter white dwarfs of
comparable mass). However, the relatively small derived values for the
stellar mass and the total envelope mass remain a concern, implying that
double-layered models may be less able to explain the observations of both
GD~358 and CBS~114 in a physically self-consistent manner
\citep[cf.][]{met03}. Additional calculations of the mass-dependence of
time-dependent diffusion profiles in white dwarf envelopes and higher
signal-to-noise observations of CBS~114 to yield a larger number of
periods for fitting would both clearly be useful.

Comparison of our model to that of \citeauthor{fb02} revealed a systematic
difference of 1,900 K between our effective temperature scales. This may
help us to understand the relatively low effective temperatures derived in
previous studies using our models \citep{mwc01,msw02,met03}. The source of
this temperature offset appears to be related to our use of the larger LAO
radiative opacities, and possibly due to subtle differences between our
convective prescriptions and the equations of state that we use for both
the core and the envelope.

Our understanding of stellar interiors through white dwarf
asteroseismology is evolving rapidly. It is now clear that the pulsations
really do sample the star globally, and we must be careful to avoid the
potential ambiguities caused by the intrinsic core/envelope symmetry. Our
challenge is to find a physically self-consistent description of both the
core and the envelope that will allow us to match the pulsation periods
within the observational uncertainties. With persistence, and with an
open-source development philosophy for our models \citep{met02}, the
future holds great promise for unveiling the detailed composition and
structure of white dwarf interiors.

\vspace*{-12pt}
\section*{ACKNOWLEDGMENTS}

We would like to thank Don Winget for helpful discussions. This research
was supported in part by the Smithsonian Institution through a CfA
Postdoctoral Fellowship, and by the UK Particle Physics and Astronomy
Research Council. S.D.K. appreciates support from the NASA Astrophysics
Theory Program through grant number NAG5-8352 to Iowa State University.
Computational resources were provided by White Dwarf Research
Corporation\footnotemark[2].

  \footnotetext[2]{http://WhiteDwarf.org}

\end{document}